\documentclass[final,1p,times]{elsarticle} 
\usepackage{graphicx} 
\usepackage{amssymb} 
\usepackage{amsthm} 
\usepackage{lineno} 

\journal{Nuclear Physics A} 

\def\ie{{\sl i.e.\/}}
\def\etal{{\sl et al.\/}}
\def\L{\langle L\rangle}
\def\ppbar{\overline\psi\psi}
\def\jhep{{\sl J.\ H.\ E.\ P.\/}}
\def\np{{\sl Nucl.\ Phys.\/}}
\def\pl{{\sl Phys.\ Lett.\/}}
\def\pr{{\sl Phys.\ Rev.\/}}
\def\prl{{\sl Phys.\ Rev.\ Lett.\/}}
\begin{document} 

\begin{frontmatter} 


\title{Exploring the gluo$N_c$ plasma}

\author{Saumen Datta and Sourendu Gupta}

\address{Department of Theoretical Physics, Tata Institute of Fundamental Research, Homi Bhabha Road, Mumbai 400005, India}

\begin{abstract} 
We report lattice computations in SU($N_c$) pure gauge theory, where $N_c$
is increased beyond the physical value of 3.  We demonstrate two-loop
scaling of $T_c$, thus obtaining the variation of $T_c/\Lambda{\overline
{\scriptscriptstyle MS}}$ with $N_c$, and fixing the temperature scale. We
study the equation of state of the gluo$N_c$ plasma, the conformal
anomaly, and the approach to the weak coupling theory. We find that the
weak-coupling prediction is always closer to the lattice data than the
conformal field theory is.
\end{abstract} 

\end{frontmatter} 



\section{The phase diagram at large $N_c$}
\label{sc.phase}

\begin{figure}[hbt]
\begin{center}
   \scalebox{0.45}{\includegraphics{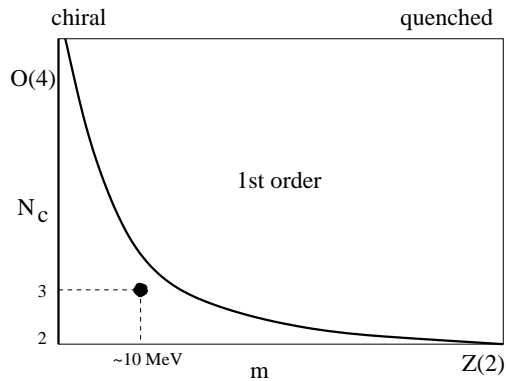}}
\end{center}
\caption{The flag diagram of QCD in the plane of $N_c$ and the quark
  mass $m$, with two flavours of degenrate quarks, \ie, with a flavour SU(2)
  group. For physical values of $N_c$ and $m$ there is only a cross over
  at finite temperature. For each fixed value of the quark mass, $m$, there
  is a critical end point of the first order deconfining line, $N_c^*$. As
  $m$ changes, this end point changes continuously, going away to infinity
  in the chiral limit $m\to0$. This line of the critical end point of the
  deconfinement transition is in the Ising universality class. The line
  $m=0$ is a critical line for the chiral phase transition and is expected
  to lie in the O(4) universality class.}
\label{fg.phase}
\end{figure}

The phase diagram of large $N_c$ QCD can be patched together from two
different universality arguments. The first is about the pure
gauge theory \cite{sveyaf}. The order parameter for the finite temperature
($T$) transition in this theory is the expectation value of the Wilson
line, $\L$.  This vanishes at small $T$, but could be non-vanishing at
high $T$ \cite{polyakov}. The loop is unchanged under the center symmetry of
the gauge group, Z($N_c$). Therefore, if there is a phase transition,
then it lies in the universality class of Z($N_c$). Since there is no
critical point for any $N_c>2$, such transitions, if they occur, must
be of first order. The SU(2) theory is an exception, and would have a
second order transition. In other words, we might expect a line of first
order transitions ending in a critical point at $N_c=2$ which is in the
Z(2) universality class.  There is evidence for such a line from many
studies with $N_c=3$ \cite{nc3}, and several with $N_c=4$ \cite{nc4}
and higher \cite{largen}, ending with a critical point in the correct
universality class \cite{nc2}.

Since quark loops can be neglected to an accuracy of $1/N_c$ in QCD
at large $N_c$, the quark determinant in the QCD partition function is
suppressed by a power of $N_c$, and hence can be neglected.  As long as
the quark mass, $m$, is large enough to ensure this, continuity arguments
(\ie, the Gibbs' phase rule) ensures that one should find a line of
first order phase transitions ending in a critical point in the Ising
universality class.  In other words, the $N_c=2$ critical point in the
quenched theory develops into an Ising critical line with changing quark
mass mass. This critical line bounds a region of first order transitions.

In the chiral limit, on the other hand, the effect of the quark
determinant on thermodynamics cannot be neglected. Here one expects chiral
symmetry breaking for two flavours of massless quarks, and hence three
massless pions at low temperature. This leads to a second universality
argument \cite{piswil}. For generic $N_c$ one might expect that chiral
symmetry is restored at finite temperature, the order parameter being the
chiral condensate, $\ppbar$. Since SU(2) flavour symmetry restoration is
in the same universality class as the O(4) spin model, we expect a line
of second order phase transitions in this universality class for generic
$N_c$. For $N_c=2$, since all representations are real, the symmetry is
enhanced to the Pauli-G\"uersey symmetry, U(4), spontaneously broken to
Sp(4). The finite temperature symmetry restoring transition is critical,
and in a different universality class, possibly O(6) \cite{nc2nf2}.

The quark determinant adds terms of the kind $(\log\lambda_i)/N_c$
to the QCD action, where $\{\lambda_i\}$ are eigenvalues of the Dirac
operator. As $m$ decreases, the lowest eigenvalue, $\lambda_0$, moves to
zero linearly with $m$, and a finite density of eigenvalues develops near
$\lambda_0$. This mechanism is intimately related to chiral symmetry
breaking \cite{banks}. Then, as $m$ decreases, one must move to larger
and large $N_c$ in order to be able to apply the large $N_c$ arguments.
For $N_c=3$ and roughly physical quark masses one has a thermal crossover,
so this ``physical'' point lies outside the boundary where the large
$N_c$ arguments about the phase diagram apply.

What can one learn about QCD from a study of the large $N_c$ pure
gauge theory?  It would seem that there is little that one can learn
quantitatively about the phase diagram of interest, \ie, for QCD
with light quark masses, since the pure gauge theory has a first order
deconfinement transition whereas that theory has a cross ov er.  However,
the high temperature phase over the full flag diagram is connected. As a
result, one thing that one might learn about this more physically relevant
theory is the nature of the high temperature phase. In particular we
ask questions about the approach to the ideal gas, and the relevance
of strongly coupled conformal theories and the weak coupling expansion
in this phase. Such questions are likely to have qualitatively similiar
answers in every part of the flag diagram.

\section{The temperature scale}
\label{sc.scale}

\begin{figure}[tbh]
\begin{center}
   \scalebox{0.55}{\includegraphics{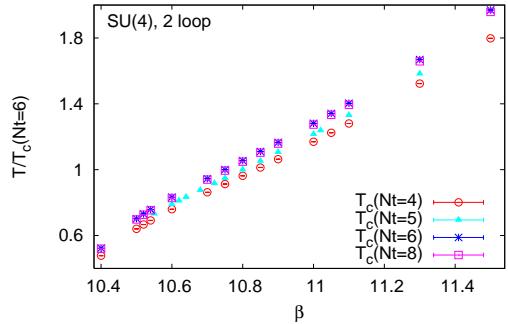}}
\end{center}
\caption{The temperature scale in the V-scheme set using two-loop RGE for
  SU(4) pure gauge theory. The lattice spacing at $T_c$ is small enough
  for two-loop scaling to be good for $N_t=6$ or more.}
\label{fg.scale}
\end{figure}

The computations reported here were performed with $N_c=4$ and 6 on
$N_t\times N_s^3$ lattices with $N_t=4$, 6 and 8, and $N_s\ge2N_t$. These
include the finest lattice spacings used in simulations in four
dimensions for $N_c>3$ uptil now. The Wilson action was used to study the
deconfinement transition temperature, $T_c$.  In the Euclidean thermal
field theory $N_t a(g) = 1/T$, where $a(g)$ is the lattice spacing when
the gauge coupling is $g$. By making an independent measurement of $g$
and then using the renormalization group equation (RGE) for finding
$a(g)$, one can set the scale of $T$. The RGE needs a boundary condition,
which is supplied by the determination of $T_c$. This process gives the
temperature scale $T/T_c$ \cite{precise}.

The renormalized coupling was defined by a measurement of the $T=0$
plaquette.  The weak coupling expansion of the plaquette to second
order was inverted to obtain the running coupling at the scale of the
plaquette, which is proportional to $a$. This is called the V-scheme
\cite{vscheme}. Then the two-loop RGE was solved to give $T/T_c$. The
adequacy of the two-loop RGE can be gauged by comparing the scales
obtained by simulations with different $N_t$. Figure \ref{fg.scale}
shows that two-loop scaling is adequate for $N_t=6$ or more when $N_c=4$.

The scale setting of $T/T_c$ also allows us to invert the RGE and
find $T_c/\Lambda_{\overline{MS}}$, in other words, to determine
$\Lambda_{\overline{MS}}$. This is done in the V-scheme for each $N_c$.
The result scales to a finite value in the $N_c\to\infty$ limit with
$1/N_c$ corrections.  The determination of $\Lambda_{\overline{MS}}$
could be more sensitive to the scheme choice than the temperature scale
because of the logarithmic running of the coupling.

\section{The equation of state}
\label{sc.eos}

\begin{figure}[htb]
\begin{center}
   \scalebox{0.5}{\includegraphics{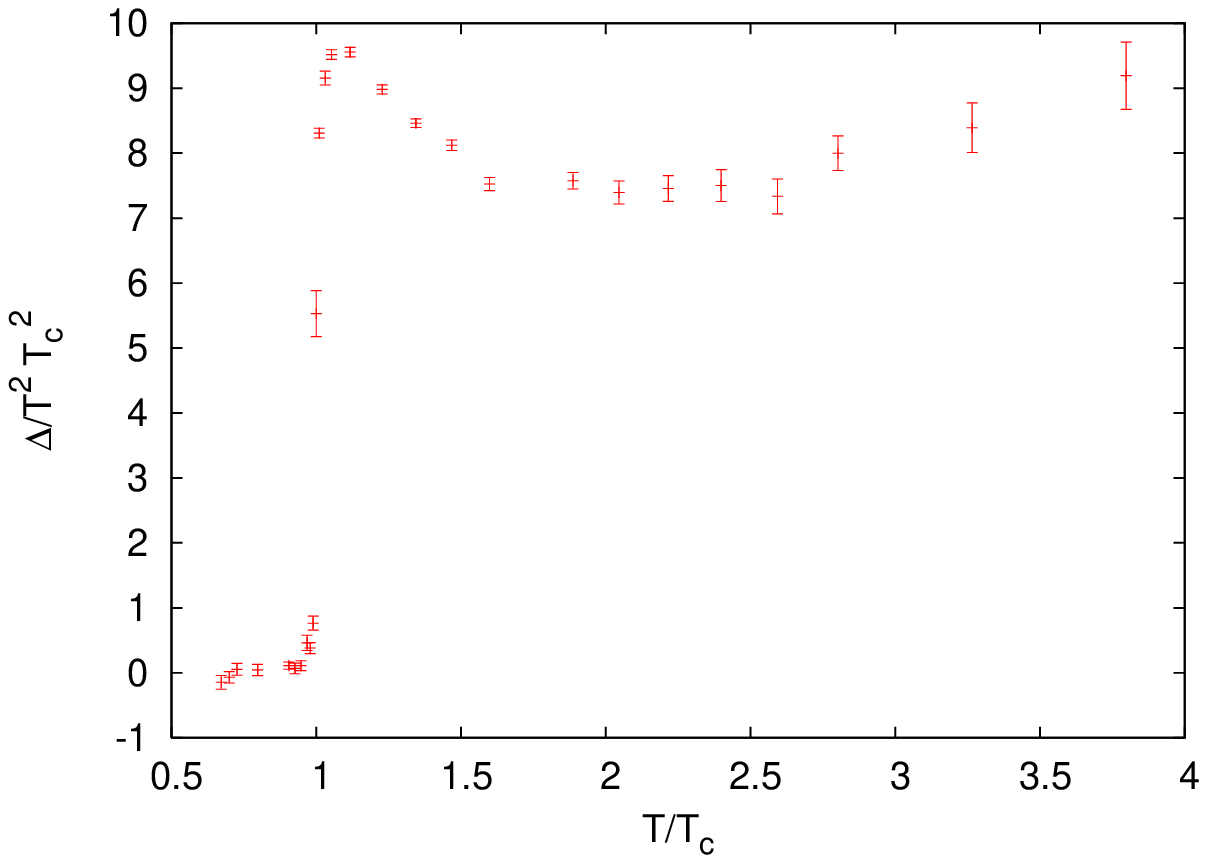}}
   \scalebox{0.5}{\includegraphics{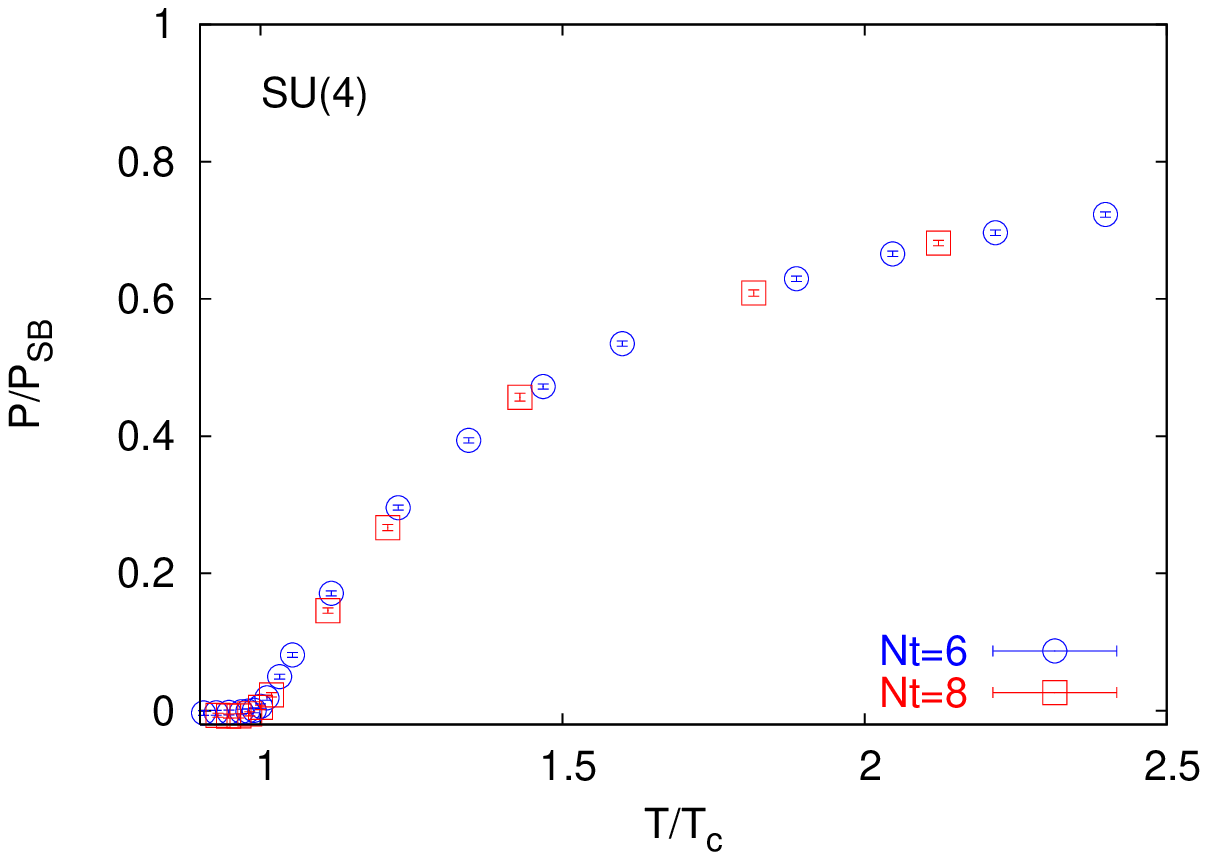}}
\end{center}
\caption{The equation of state for SU(4) pure gauge theory. On the left
  is the interaction measure $\Delta/(T^2T_c^2)$, showing that it scales
  as $T^2$. On the right is the pressure, $P$, evaluated at two different
  lattice spacings, showing that the continuum result is obtained.}
\label{fg.eos}
\end{figure}

The equation of state was computed for SU(4) pure gauge theory.
The interaction measure, $\Delta=E-3P$, computed using the two-loop beta
function, scales as $T^2$ \cite{pisarski} in the high-temperature phase
(see Figure \ref{fg.eos}). Computing the pressure using the integral
method, we found that the continuum limit is reached at $N_t=6$.

\begin{figure}[hbt]
\begin{center}
   \scalebox{0.5}{\includegraphics{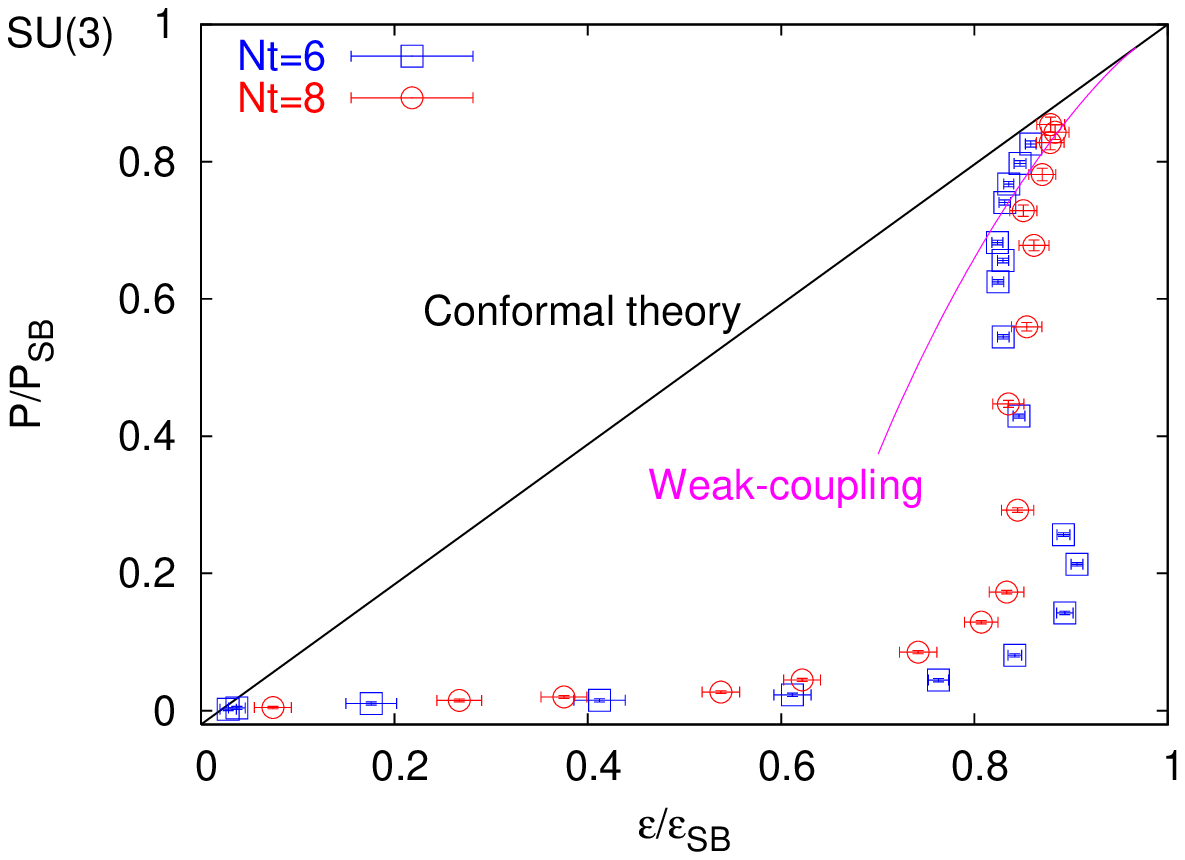}}
   \scalebox{0.5}{\includegraphics{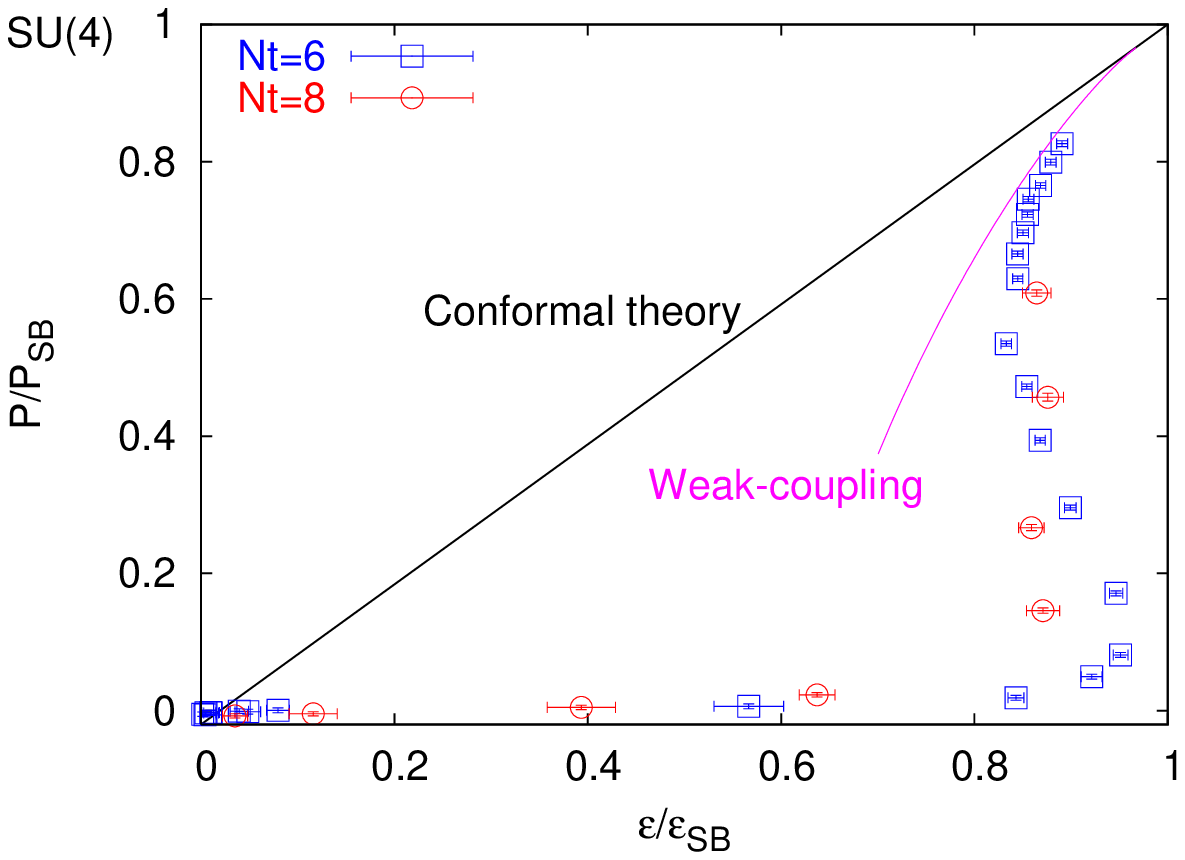}}
\end{center}
\caption{The equation state of SU(3) (left) and SU(4) (right) pure
  gauge theory. The diagonal is the line of all possible conformal
  theories ($\Delta=0$), and the curve is the weak coupling resummed
  result from \cite{mikko}.}
\label{fg.confo}
\end{figure}

We show the equation of state of the SU(3) \cite{nc3} and SU(4) theories
in Figure \ref{fg.confo}. We see that the lattice data are closer to
the weak-coupling results \cite{mikko} than to conformal field theory
($\Delta=0$) at all temperatures.  At high temperatures, of course,
the weak coupling theory tends to an ideal gas, which in turn is a
conformal theory.  There seems to be no window in temperature where a
non-trivial conformal theory describes the equation of state better than
weak-coupling theory.

\end{document}